\documentclass[aps,prx,twocolumn,english,superscriptaddress,floatfix,longbibliography]{revtex4-2}

\usepackage[english]{babel} 
\usepackage[T1]{fontenc} 
\usepackage[utf8]{inputenc} 
\usepackage{graphicx} 
\usepackage{amsmath}
\usepackage{amssymb}
\usepackage{comment}
\usepackage{units}
\usepackage{upgreek}
\usepackage{braket}
\usepackage{natbib}
\usepackage[colorlinks=true,bookmarks=false,citecolor=blue,urlcolor=blue]{hyperref}
\usepackage{lipsum}
\usepackage{orcidlink}
\newcommand*{\figref}[2][]{%
  Fig.~\hyperref[{fig:#2}]{%
    \ref*{fig:#2}%
    \ifx\\(#1)\\%
    \else
      (#1)%
    \fi
  }%
}
\renewcommand{\selectlanguage}[1]{}
\begin{document}
\title{Heralded initialization of charge state and optical transition frequency of diamond tin-vacancy centers}
\author{Julia M. Brevoord\orcidlink{0000-0002-8801-9616}}
\thanks{These two authors contributed equally}
\author{Lorenzo De Santis\orcidlink{0000-0003-0179-3412}}
\thanks{These two authors contributed equally}
\author{Takashi Yamamoto\orcidlink{0009-0004-8038-4303}}
\author{Matteo Pasini\orcidlink{0009-0005-1358-7896}}
\author{Nina Codreanu\orcidlink{0009-0006-6646-8396}}
\author{Tim Turan\orcidlink{0009-0003-9908-7985}}
\author{Hans K. C. Beukers\orcidlink{0000-0001-9934-1099}}
\author{Christopher Waas\orcidlink{0009-0008-1878-2051}}
\author{Ronald Hanson\orcidlink{0000-0001-8938-2137}}
\email{R.Hanson@tudelft.nl}
\affiliation{QuTech and Kavli Institute of Nanoscience$,$ Delft University of Technology$,$ Delft 2628 CJ$,$ Netherlands}
\date{\today}

\begin{abstract}
Diamond Tin-Vacancy centers have emerged as a promising platform for quantum information science and technology. A key challenge for their use in more complex quantum experiments and scalable applications is the ability to prepare the center in the desired charge state with the optical transition at a pre-defined frequency. Here we report on heralding such successful preparation using a combination of laser excitation, photon detection, and real-time logic. We first show that fluorescence photon counts collected during an optimized resonant probe pulse strongly correlate with the subsequent charge state and optical transition frequency, enabling real-time heralding of the desired state through threshold photon counting. We then implement and apply this heralding technique to photoluminescence excitation measurements, coherent optical driving, and an optical Ramsey experiment, finding strongly improved optical coherence with increasing threshold. Finally, we demonstrate that the prepared optical frequency follows the probe laser across the inhomogeneous linewidth, enabling tuning of the transition frequency over multiple homogeneous linewidths.
\end{abstract}

\maketitle

\section{Introduction} 
In the past decade, color centers in diamond have become a leading platform for quantum networking experiments~\cite{atature_material_2018,ruf_quantum_2021,wolfowicz_quantum_2021}. 
The first demonstrations relied on the Nitrogen-Vacancy (NV) center, ranging from the first loophole-free Bell test~\cite{hensen_loophole-free_2015} to a multi-node network experiments~\cite{pompili_realization_2021,hermans_qubit_2022}. However, the NV center suffers from a low Debye-Waller factor and strong spectral diffusion close to surfaces, making its optical interface inefficient. Group-IV vacancy centers emerged as a favourable alternative due to their high Debye-Waller factor~\cite{iwasaki_tin-vacancy_2017,bradac_quantum_2019, ruf_quantum_2021} and inversion symmetry~\cite{hepp_electronic_2014}, resulting in first-order insensitivity to charge noise~\cite{de_santis_investigation_2021} and thus compatibility with nanophotonics integration~\cite{sipahigil_integrated_2016,evans_photon-mediated_2018}. Pioneering experiments showing basic network node operation in a dilution refrigerator have been performed using the silicon-vacancy (SiV) center in nanophotonic devices~\cite{sukachev_silicon-vacancy_2017,bhaskar_experimental_2020,stas_robust_2022}.
More recently, the negatively charged Tin-Vacancy (SnV$^-$) center in diamond has attracted significant interest thanks to a high quantum efficiency~\cite{iwasaki_tin-vacancy_2017, sipahigil_integrated_2016,nguyen_photodynamics_2019} and significant spin-orbit coupling, which allows for operation at elevated temperatures compared to the SiV center~\cite{iwasaki_tin-vacancy_2017,gorlitz_spectroscopic_2020,trusheim_transform-limited_2020}. In recent experiments, the integration of SnV centers into nanophotonic devices~\cite{rugar_quantum_2021, arjona_martinez_photonic_2022, kuruma_coupling_2021, pasini_nonlinear_2023} and coherent SnV spin qubit control have been demonstrated~\cite{rosenthal_microwave_2023,guo_microwave-based_2023}.\\
\\
It has been widely reported that under resonant optical excitation the SnV$^-$ center can go into a dark state~\cite{iwasaki_tin-vacancy_2017,gorlitz_spectroscopic_2020,trusheim_transform-limited_2020,arjona_martinez_photonic_2022}, possibly the SnV$^{2-}$ state~\cite{gorlitz_coherence_2022}. Off-resonant excitation (typically around \unit[515-532]{nm}) is routinely used to repump to the SnV$^{-}$ state~\cite{gorlitz_spectroscopic_2020, rugar_quantum_2021, de_santis_investigation_2021,arjona_martinez_photonic_2022}. The optical transition frequency of the SnV$^{-}$ after the repump is in general not the same as before (spectral diffusion), possibly caused by the capture or release of nearby charges in the repump process or by the direct impact of the pump excitation on the charge environment~\cite{gorlitz_coherence_2022}. These processes pose two challenges for future use in quantum protocols: i) the charge state repump is probabilistic, leading to initialization errors, and ii) spectral diffusion hinders efficient optical spin initialization and readout and causes reduced photon indistinguishability, which negatively impacts key protocols requiring photon interference such as remote entanglement generation~\cite{beukers_tutorial_2023}. In this work, we overcome these challenges by realizing heralded initialization into the desired charge state and pre-set the optical transition frequency. 

\section{Experimental setup}
A simplified level structure of the SnV center in the negative charge state (SnV$^{-}$) is shown in~\figref[a]{fig1}. In the absence of a magnetic field, the ground and optically excited states consist of spin-degenerate orbital doublets, leading to four optical transitions. Of main interest for a spin-photon interface is the Zero-Phonon Line (ZPL) transition at \unit[619]{nm}, linking the lowest-branch ground and optically excited state. Besides the ZPL path, photon emission can be accompanied by the excitation of vibronic modes (phonon-side band emission, PSB).\\
\\
Our main investigation focuses on an SnV$^{-}$ center (labeled SnV$^-$A) in an IIa <111>-oriented diamond, where Sn-ions are implanted $\sim$\unit[1]{nm} below the surface. Subsequently, overgrowth of the diamond by chemical vapor deposition~\cite{rugar_generation_2020} results in SnV$^{-}$ centers $\sim$\unit[550]{nm} below the surface. All experiments on SnV$^-$A are performed in an optical confocal set-up at \unit[4]{K} (see Supplemental Material~\cite{Supplemental_Material} for experimental set-up details, which includes~\cite{raa_qmi_2023,brevoord_data_2023}). \figref[b]{fig1} shows the second-order correlation function, $g^{(2)}$, of the PSB emission of the SnV$^{-}$ center under continuous resonant excitation. We find $g^{(2)}(0) = 0.10\pm0.02$ without any background subtraction showing that the light indeed originates from an individual emitter.
\begin{figure}
	\includegraphics[width=\linewidth]{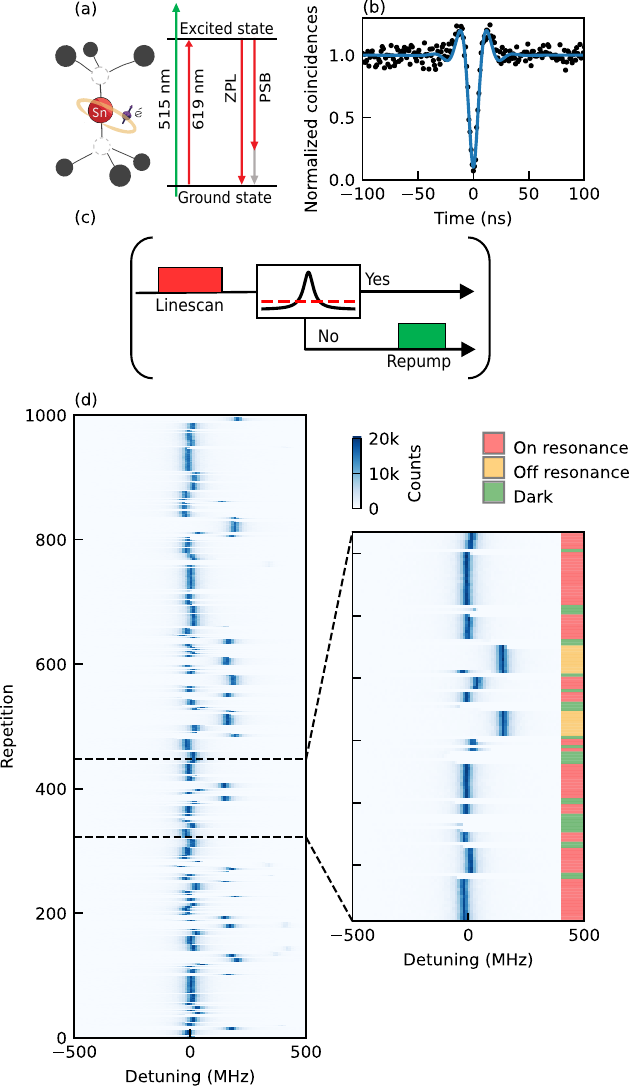}
	\caption{(a) Schematic of the color center structure and the optical transition of interest of an SnV$^-$ center in diamond. Resonant excitation at \unit[619]{nm} results in emission in the ZPL or PSB decay channels. (b) Second-order correlation measurement of the emission from a single SnV$^-$ under continuous resonant excitation of \unit[80]{nW}, fitted by $g^{(2)}(\uptau)=1-e^{\frac{-3\uptau \upgamma}{4}}(\cos(\upomega \uptau)+\frac{3\upgamma}{4\upomega}\sin(\upomega \uptau))$, obtaining a $g^{(2)}(0)=0.10\pm0.02$, without background correction. (c) Pulse sequence for the conditional PLE measurements, where we apply a repump pulse in case the maximum number of photon counts detected during a single frequency step is less than 1.5 times the mean number of photons detected. (d) The fluorescence of 1000 PLE scans taken in $\sim$\unit[1.3]{GHz/s} over the optical transition. The state of the emitter is estimated from the counts detected during each scan. If the emitter is flagged as dark, an off-resonant \unit[515]{nm} pulse is applied before the following scan. } \label{fig:fig1}
\end{figure}
\section{Dynamics of charge state and resonance frequency}
We investigate the SnV$^{-}$ center by performing photoluminescence excitation (PLE) scans, where a resonant laser of \unit[3]{nW} is scanned at $\sim$\unit[1.3]{GHz/s} over the optical transition while recording the PSB emission. In this experiment a conditional repump is used: we apply a repump pulse in case the peak photon counts detected during the PLE is below a pre-set threshold, see pulse sequence in~\figref[c]{fig1}. \\
\\
\figref[d]{fig1} shows 1000 consecutive resonant scans. We identify three different regimes, that we label by color on the right for a subset of scans. Red indicates that the optical transition is found near zero detuning ('on resonance') and the photon count threshold is met. This condition constitutes the desired state. Green indicates scans with counts below the threshold, indicating the emitter has gone into the dark state. A \unit[515]{nm} repump pulse of \unit[50]{ms} and \unit[1]{$\upmu$W} is applied prior to the following scan. It can be seen that the repump pulse indeed brings the emitter back to the bright SnV$^{-}$-charge state with good probability, often accompanied by a shift of the resonant frequency of the emitter. We highlight scans with orange in which the threshold of counts is met, but the detected resonance is significantly detuned, >\unit[100]{MHz} ('off resonance'). Importantly, we find that the SnV$^{-}$ exhibits high spectral stability both in the on-resonance and off-resonance conditions, up to the point that ionization occurs.

\section{Probe pulses for charge state and resonance condition detection}
Motivated by the observed spectral stability before ionization, we explore the possibility of using photon counts during a resonant probe pulse as a heralding signal for the successful preparation of the SnV center in the negative charge state with its optical transition at a desired frequency.
To gain quantitative insights into the predicting capabilities of such heralding signal, we implement the pulse sequence shown in~\figref[a]{fig2}. A \unit[515]{nm} 'repump' pulse is followed by two identical resonant laser pulses, named here 'Pulse 1' and 'Pulse 2'. The sequence in~\figref[a]{fig2} is repeated $10,000$ times in our experimental runs.
\\
In~\figref[b]{fig2} we plot in log-scale the distributions of the photons detected during Pulse 2, $C_2$, as a function of the number of photons detected during the preceding Pulse 1, $C_1$, for three different resonant laser powers. For the lowest power (top panel), we find an almost perfect correlation between $C_1$ and $C_2$. This observation implies that the number of photons scattered by the SnV$^{-}$ center during these pulses is dictated by their charge state and their instantaneous detuning from the laser frequency. By increasing the resonant laser power (middle and lower panel), we observe a change in the distribution of photon counts. As expected, the mean number of photon counts is increased. In addition, while the correlation between $C_1$ and $C_2$ is still present, we can see the effect of ionization, resulting in vertical and horizontal regions of uncorrelated photon distributions. The vertical band is due to ionization during Pulse 2 after several photons have already been detected. The horizontal band mainly corresponds to cases where ionization occurred during Pulse 1; those cases could lead to an incorrect heralding signal and should thus be minimized.\\
\begin{figure}
	\includegraphics[width=\linewidth]{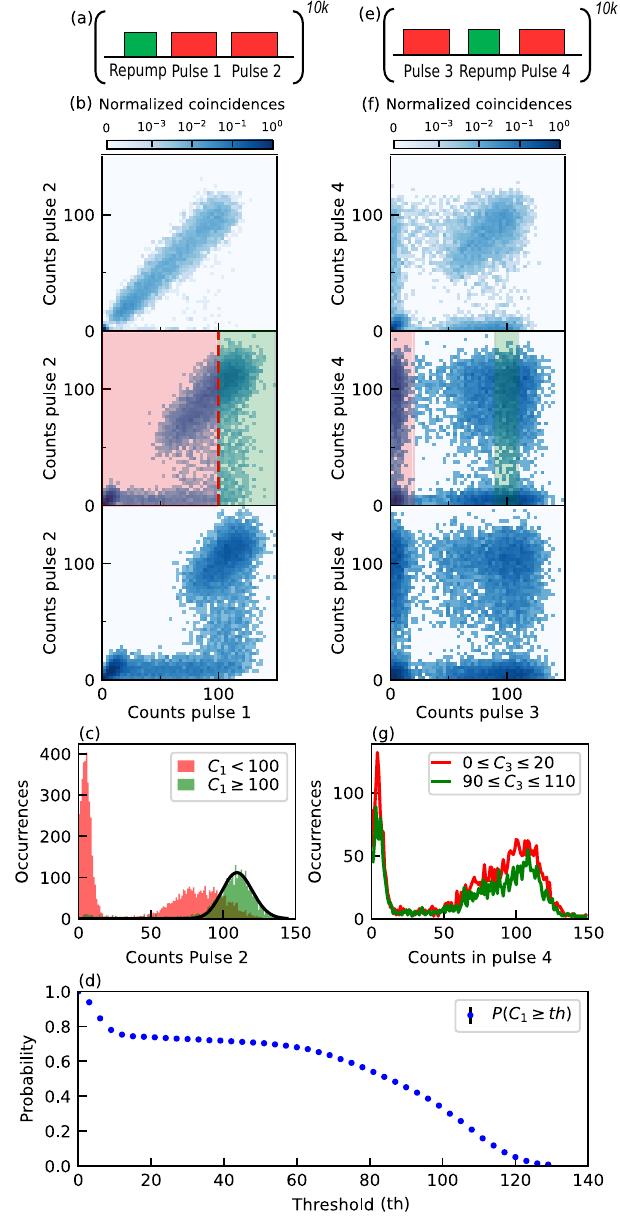}
	\caption{(a) The pulse sequence used for the data in (b), consisting of an off-resonant pump pulse followed by two identical resonant probe pulses (1 and 2). (b) 2D histogram of photon counts $C_2$, as a function of $C_1$ for increasing resonant laser powers (\unit[10]{nW}, \unit[100]{nW}, \unit[200]{nW}) and a fixed duration of \unit[50]{$\upmu$s}. (c) Histogram of $C_2$ conditioned on $C_1$ above or below a threshold of 100 counts of the data of the middle panel of (b). (d) The probability of passing the threshold used for heralded bright state initialization, the error bars fall within the data points. (e) The pulse sequence used for the data in (f), consists of two identical resonant probe pulses (3 and 4) with an off-resonant pump pulse in between. (f) 2D histogram of photon counts $C_4$, as a function of $C_3$ for increasing repump duration and powers (\unit[50]{$\upmu$s} at \unit[10]{$\upmu$W}, \unit[500]{$\upmu$s} at \unit[100]{$\upmu$W}, \unit[750]{$\upmu$s} at \unit[200]{$\upmu$W}). (g) Histogram of $C_4$ conditioned on $C_3$ being a low or high number of counts. 
\label{fig:fig2}}
\end{figure}\\
Having confirmed that the probe signal strongly correlates to the state of the SnV$^{-}$ center after the probe, we now use this to condition our data set to cases where the SnV$^{-}$ is on resonance with the probe laser (i.e. a high number of photon counts in the detection window). As shown in~\figref[c]{fig2}, if we condition on $C_1 >100$, the photon count distribution of Pulse 2 follows a Poisson distribution centered around 110 counts, while the reverse conditioning shows a broader, lower-counts distribution mixed with a peak near zero counts. The thresholding approach demonstrated here allows to filter the desired bright state condition of the color center out of the statistical distribution of possible charge-resonance states. Tighter thresholding can be done at the expense of a decrease in efficiency, as shown in~\figref[d]{fig2}. For instance, for the threshold value of 100 used above, we observe a success probability of passing the threshold of 33\%. \\
\\
Similarly to the analysis above, we study the effect of the repump laser (see pulse sequence in~\figref[e]{fig2}) by plotting the number of photons detected during Pulse 4, $C_4$, as a function of the number of photons detected in the probe pulse ($C_3$) preceding a pump pulse, see~\figref[f]{fig2}. For weak repump pulses (top panel), some correlation between the probe and pump signals is visible, as the repump in this case does not significantly affect the SnV center and its environment. For sufficiently strong repump pulses (middle and lower panel), no correlations are observed, showing that the state following the repump pulse is independent of the state before the repump. This is confirmed by the histograms shown in~\figref[f]{fig2}, where we plot the counts distribution of $C_4$ of the data of the middle panel of~\figref[f]{fig2} for the 2 shaded areas. We see that the distributions are similar for low and high counts detected in Pulse 3.\\
\\
Using again the data of the middle panel of~\figref[g]{fig2}, we estimate how efficiently the repump re-initializes the SnV$^{-}$ in the negative charge state if it was in the dark state before. For this, we set a threshold of 20 counts to distinguish between the desired charge state and the dark state. Conditioning on a dark state being detected on Pulse 3, the probability of detecting a bright state on Pulse 4 reaches about 75$\%$.

\section{Optical Rabi driving}
Next, we investigate the correlation between probe pulse counts and the SnV$^{-}$ optical coherence. We apply the pulse sequence as shown in~\figref[a]{fig3}, where the repump and probe pulse are now followed by 500 repetitions of a \unit[30]{ns} resonant pulse used to drive optical Rabi oscillations. \figref[b]{fig3} shows a histogram of the number of photons detected during the probe pulse, where 3 peaks are present. We allocate the instances of high detected photon counts (rightmost peak) to the emitter being in the 'on resonance' state. The instances of the middle and left peak instead correspond respectively to the 'off-resonant' and 'dark' cases. Based on this, we divide the histogram into four parts, as highlighted by the shaded background colors of~\figref[b]{fig3} corresponding to the 3 peaks described above plus an intermediate region between the center and rightmost peak.\\
\\
\figref[c]{fig3} shows time traces of the detected photons during the \unit[30]{ns} resonant readout pulse conditioned on the four threshold intervals. Each curve reveals coherent driven oscillations with different amplitude, frequency, and decay times. As expected, by thresholding for higher probe counts we obtain higher count rates in the readout, as we are selecting for cases where the SnV$^{-}$ is on resonance with the driving laser. By fitting these curves with an exponentially damped sine function we extract the Rabi frequency and decay time of the oscillations. These values are summarized in~\figref[d]{fig3}, showing that the fitted decay time (frequency) increases (decreases), consistent with the SnV$^{-}$ being closer to resonance for an increasing number of detected photons in the probe pulse. These results demonstrate a clear relation between the probe counts and the measured coherence.
\begin{figure}
	\includegraphics[width=\linewidth]{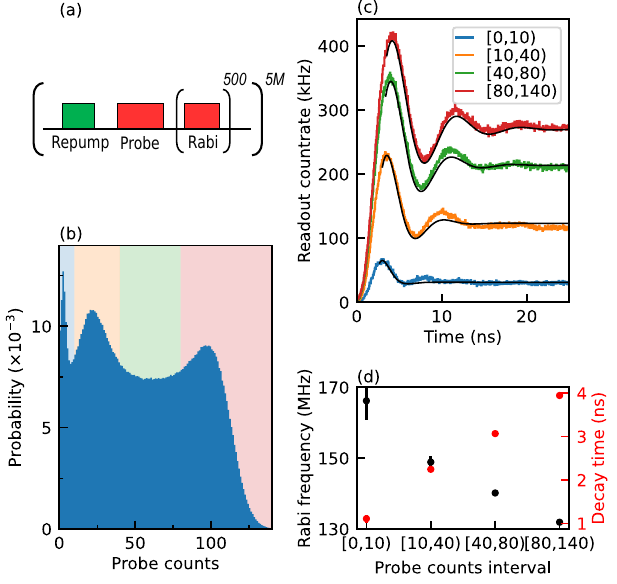}
	\caption{(a) The pulse sequence used for the Rabi driving experiment conditioned on the probe pulse counts (Repump: \unit[500]{$\upmu$s}, \unit[100]{$\upmu$W}, Probe: \unit[500]{$\upmu$s}, \unit[100]{nW}, Rabi pulse: \unit[30]{ns} at \unit[17.5]{nW}). (b) Histogram of the photon counts detected during the probe pulse. (c) Time-resolved histograms of photon counts during Rabi driving, conditioned on the number of photons detected in the preceding probe pulse. Solid lines are exponential decaying sine fits to the data. (d) The decay times and Rabi frequencies were obtained by fitting the time traces in panel (c), including the error of the fits.
\label{fig:fig3}}
\end{figure}

\section{Real-time heralding of charge state and optical transition frequency }
In the experiments so far, the conditioning on probe pulse counts was done in post-selection. For scalable applications in quantum protocols, it is key that the selection is done in real-time i.e. before quantum protocols are run~\cite{pompili_realization_2021,hermans_entangling_2023, hermans_qubit_2022, stolk_telecom-band_2022}. In the following, we implement live thresholding on the probe counts using the programmable logic of a fast microcontroller (running on a \unit[10]{$\upmu$s} clock cycle, see Supplemental Material~\cite{Supplemental_Material}) and use this as a charge resonance check (CRC) routine to herald the desired charge-resonance state before each experimental run.\\
\\
The CRC sequence (see ~\figref[a]{fig4}) starts with a resonant probe pulse and live counting. Below we report on an implementation using two threshold values instead of just one, which allows more freedom in trading off heralding efficiency (rate) and accuracy. In the case that the number of counts detected during the resonant probe pulse is below $C_{\text{repump}}$, an off-resonant \unit[515]{nm} repump pulse is applied, followed by a resonant probe pulse. In case the counts detected during the probe pulse are above $C_{\text{repump}}$ but below $C_{\text{pass}}$, the resonant probe pulse is applied again. $C_{\text{repump}}$ probes whether the emitter is in the correct charge state and the threshold $C_{\text{pass}}$ functions to filter for instances that the emitter is not on resonance with the driving field. This procedure is repeated until the threshold $C_{\text{pass}}$ is met. \\
\\
We first implement the CRC in conjunction with PLE scans to show its effect on spectral diffusion and ionization, see~\figref[b]{fig4}. The CRC repump and probe pulses are of the same power and duration as in~Fig.~\ref{fig:fig3}. \figref[b]{fig4} depicts two panels with 250 PLE scans each. Before every scan, a CRC is performed with a ($C_{\text{pass}}$, $C_{\text{repump}}$) threshold of (50, 10) and (110,10) counts for the left and right panel respectively.\\
\\
For a low CRC threshold, $C_{\text{pass}}=50$, the SnV$^{-}$ resonant frequency shows spectral jumps less frequently compared to a high CRC threshold, $C_{\text{pass}}=110$, but jumps with higher magnitude. Due to the lower threshold, enough photons can be scattered to pass the CRC even when the SnV$^{-}$ is off-resonance. As a result, these off-resonant cases affect the PLE experiment, while the repump pulse is applied only once the emitter is dark. \\
\begin{figure}
	\includegraphics[width=\linewidth]{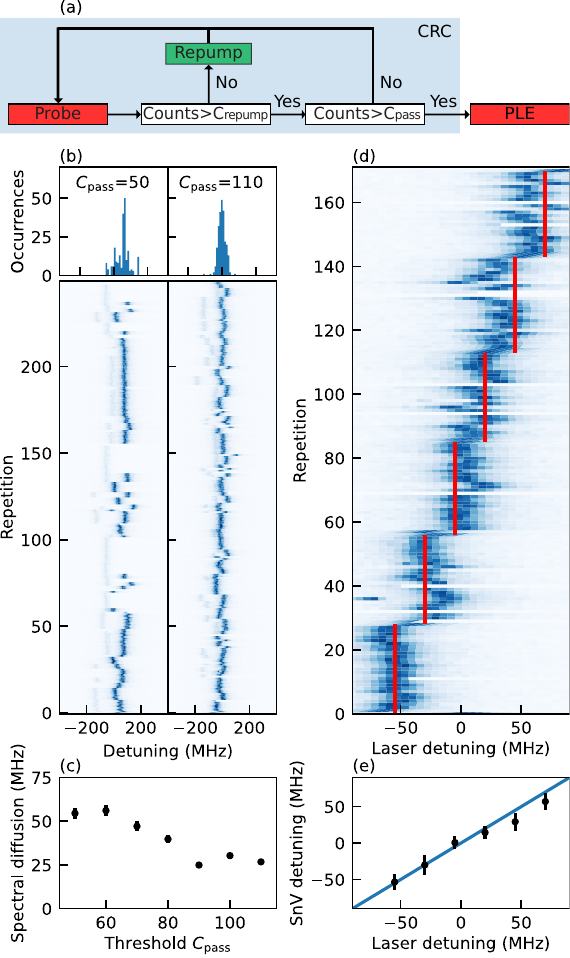}
	\caption{(a) The real-time logic pulse sequence for the CRC used to herald the charge-resonance state of the SnV. (b) Bottom: The fluorescence of 300 PLE scans each taken in $\sim$\unit[1.3]{GHz/s} over the optical transition of SnV$^-$A, preceded by a CRC for a threshold $C_{\text{pass}}$ of 50 (left) and 110 (right) counts and a threshold $C_{\text{repump}}=10$. We attribute the second red-detuned resonance detected to another nearby emitter. Top: The distribution of the fitted centers of the individual scans, filtered for the bright emitter. (c) The standard deviation of the centers of the individual fitted scans as a function of the CRC threshold. The error bar is the standard error of the standard deviation. (d) The fluorescence of PLE scans each taken in $\sim$\unit[1.3]{GHz/s} over the optical transition of SnV$^-$B, preceded by a CRC for different resonant frequencies indicated by the red vertical lines. (e) The emitter’s mean central frequency as a function of the set laser frequency, showing the shift of the SnV$^-$ center’s emission by CRC conditioning. The error bar is one standard deviation over the repetitions shown in panel (d).
\label{fig:fig4}}
\end{figure}\\
With higher CRC thresholding, more repump cycles are required to reach a configuration where the SnV$^{-}$ detuning to the driving laser is small enough to scatter more photons than the threshold value, $C_{\text{pass}}$. This leads to more frequent spectral jumps of the resonance frequency but the jumps are of significantly lower magnitude.  
It can be seen how this improves the effective spectral diffusion probed by our experiment by looking at the distribution of SnV$^{-}$ resonance frequencies in the PLE scans (top panels of~\figref[b]{fig4}). Here we have filtered for the resonance of the other emitter. In~\figref[c]{fig4}, we show the standard deviation of the distribution in the top panel of~\figref[b]{fig4} as a function of the CRC threshold, $C_{\text{pass}}$. A clear trend towards a single-peaked distribution and lower variance is visible for higher thresholds (see Supplemental Material for the PLE scans~\cite{Supplemental_Material}). We note that employing a CRC threshold $C_{\text{pass}}<50$ resulted in too few repump pulses to reliably determine the repump-induced spectral diffusion. \\
\\
In addition, we applied the CRC on a different SnV (SnV$^-$B) which is embedded in a nanophotonic waveguide (we refer to~\cite{pasini_nonlinear_2023} the device and experimental setup details). When setting a high $C_{\text{pass}}$, we observe similar stable lines as for SnV$^-$A, see~\figref[d]{fig4} and Supplemental Material for more PLE scans using different CRC thresholds~\cite{Supplemental_Material}. Furthermore, we show on SnV$^-$B that the CRC can be employed to tune the heralded optical transition frequency. In~\figref[d]{fig4}, we step the frequency of the resonant laser during the CRC about every 30 scans, while using a high $C_{\text{pass}}$. The resonant peak center detected in the subsequent PLE follows the frequency set-point indicated by the red solid lines as shown in ~\figref[d]{fig4}. In~\figref[e]{fig4} we plot the mean center frequency of the individual scans for the different laser detuning setpoints. This experiment shows that using CRC heralding we can tune the emitter~>\unit[100]{MHz}, which is several times larger than the measured mean single-scan (homogeneous) linewidth of \unit[31]{MHz}. The effective tuning range of this method is determined by the inhomogeneous (or spectral-diffusion-limited) linewidth (which can be measured by setting the CRC threshold to zero), as the probability for the repump pulse to bring the optical transition to more detuned frequencies decreases rapidly.

\section{Optical Ramsey experiment using real-time heralding}
\begin{figure}
	\includegraphics[width=\linewidth]{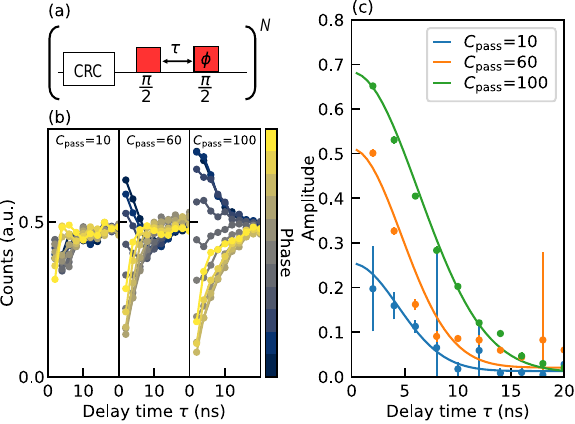}
	\caption{Optical Ramsey experiment conditioned on CRC. (a) Pulse sequence used for the data in (b) and (c) (see Supplemental Material for further details~\cite{Supplemental_Material}) (b) The counts detected during an integration window after the second $\uppi / 2$-pulse for a set $C_{\text{repump}}$ of 10 and different $C_{\text{pass}}$, for different phases of the second $\uppi / 2$-pulse, normalized to twice the counts detected in a mixed state. (c) Contrast decay envelopes for different CRC thresholds, $C_{\text{pass}}$, and a fixed $C_{\text{repump}}$ of 10. The fitted envelopes show in an increase of the $T_2^*$ from \unit[(4.3$\pm$1.7)]{ns} for a low threshold,$C_{\text{pass}}$, to $T_2^*$~=\unit[(6.3$\pm$0.4)]{ns} for the highest $C_{\text{pass}}=100$. The error bars are fit errors. \label{fig:fig5}}
\end{figure}
Finally, we directly probe the coherence of the optical transition for different CRC thresholds, by performing Ramsey interferometry experiments using the pulse sequence depicted in~\figref[a]{fig5}. First, coherence between the ground and optically excited state is created with an optical $\uppi/2$ pulse. After letting this state evolve for a time $\uptau$ we apply a second $\uppi/2$ to map the remaining coherence onto populations that we read out by integrating the fluorescence in a \unit[5]{ns} window after the second $\uppi/2$-pulse. This signal is then normalized to twice the fluorescence measured for a mixed state (i.e. for large $\uptau$). \\
\\
We run this experiment for different delays $\uptau$ and phase differences $\phi$ of the pulses. The resulting data is shown in~\figref[b]{fig5}, where the dependence of the readout signal on $\uptau$ (x-axis value) and $\phi$ (plot color) is shown for a $C_{\text{pass}}$ of 10, 60 and 110 counts in the left, center and right panel, respectively, and a fixed $C_{\text{repump}}=10$. For increasing CRC threshold $C_{\text{pass}}$ we observe both a higher contrast of the oscillations, as well as a slower decay of this contrast with increasing $\uptau$.\\
\\
We extract a quantitative measure for the coherence between the ground and excited state by fitting the phase dependence for each $\uptau$ to a sine function~\cite{arjona_martinez_photonic_2022}. The amplitude of the sine is plotted in~\figref[c]{fig5} as a function of $\uptau$. From a Gaussian fit to the data we determine the dephasing time $T_2^*$ of the optical transition. For the CRC threshold $C_{\text{pass}}=10$ we obtain a $T_2^*$ of \unit[(4.3~$\pm$~1.7)]{ns}, where the large uncertainty is a consequence of the low contrast. For a high CRC threshold of $C_{\text{pass}}=100$ we determine a $T_2^*=$ \unit[(6.3~$\pm$~0.4)]{ns}. This is, to our knowledge, the highest measured optical $T_2^*$ for an SnV$^-$ center to date. This demonstrates that implementing a CRC can mitigate the effects of spectral detuning leading to an increase in optical coherence time, which is key to improve photon interference experiments.

\section{Conclusion and Outlook}
The ability to reliably prepare color center qubits in the desired charge state at a set optical transition frequency, as demonstrated here, is a key requirement for efficiently running complex quantum experiments as well for future quantum technologies. Taken together with recent diamond SnV center demonstrations of nanophotonic integration~\cite{rugar_quantum_2021}, spin qubit control, and coherence beyond a millisecond~\cite{guo_microwave-based_2023, rosenthal_microwave_2023}, our results complete a quantum control toolkit for scaling current single-center experiments. Compared to the experimentally more mature diamond Nitrogen-Vacancy center~\cite{hermans_entangling_2023, pompili_realization_2021, stolk_telecom-band_2022, hensen_loophole-free_2015} and diamond Silicon-Vacancy center~\cite{pingault_coherent_2017, meesala_strain_2018, stas_robust_2022, bhaskar_experimental_2020}, the SnV center adds the combination of compatibility with nanophotonic integration and the relatively high quantum efficiency and operating temperature, making it a compelling platform for future exploration of multi-qubit quantum networking and quantum computing protocols.

\begin{acknowledgments}
We thank Henri Ervasti for software support. We thank Yanik Herrmann and Julius Fisher for proofreading the manuscript. We gratefully acknowledge that this work was supported by the Dutch Research Council (NWO) through the Spinoza prize 2019 (project number SPI 63-264), by the Dutch Ministry of Economic Affairs and Climate Policy (EZK) as part of the Quantum Delta NL program, by the joint research program “Modular quantum computers” by Fujitsu Limited and Delft University of Technology, co-funded by the Netherlands Enterprise Agency under project number PPS2007, and by the QIA-Phase 1 project through the European Union’s Horizon Europe research and innovation program under grant agreement No. 101102140. L.D.S. acknowledges funding from the European Union’s Horizon 2020 research and innovation program under the Marie Sklodowska-Curie grant agreement No. 840393.\\
\\

\end{acknowledgments}
\bibliography{main.bib}

\clearpage

\widetext
\begin{center}
\textbf{\large Supplemental Material: Heralded initialization of charge state and optical transition frequency of diamond tin-vacancy centers}
\end{center}
\setcounter{figure}{0}
\setcounter{page}{1}
\setcounter{section}{0}
\makeatletter
\renewcommand{\theequation}{S\arabic{equation}}
\renewcommand{\thefigure}{S\arabic{figure}}
\section{Fabrication method}\label{sec:Fabrication method}
The IIa <111>-oriented almost dislocation-free diamond substrate used for these measurements was cleaned in a piranha solution for \unit[20]{minutes} before and after a plasma etch was performed using an Ar/Cl$_2$-plasma and an O$_2$-plasma removing $\sim$\unit[8]{$\upmu$m} of the surface to relief it from polishing induced strain. The (120)Sn-ions were implanted at CuttingEdge with an energy of \unit[5]{keV} and a dose of \unit[1e11]{ions/cm$^2$}. After implantation and subsequent acid cleaning, a $\sim$\unit[550]{nm} thick diamond layer was overgrown in-house using chemical vapor deposition (CVD). During the diamond growth, the sample is brought to temperatures where the vacancies become mobile and SnV centers are formed. 
\section{Experimental set-up}\label{sec:Experimental set-up}
All measurement data in this work has been taken in a closed-cycle S50-Montana cryostat at a baseplate temperature of \unit[4]{K}. The optical set-up consists of a room-temperature confocal microscope objective that is mounted in a home-built housing with xyz-positioners (Physik Instrumente P-615K011) to allow movement of the objective. An aluminum heat shield around the room-temperature objective is used to shield the \unit[4]{K} diamond sample stage from heating by the objective. Resonant excitation around \unit[619]{nm} is performed using a Toptica TA-SHG Pro. The wavelength is stabilized by feedback through a wavemeter (HighFinesse WS/U-10U). The off-resonant \unit[515]{nm} green excitation is performed using a Hubner Photonics Cobolt 06-MLD. To create $\upmu$s resonant pulses, we use an in-fiber acousto-optic modulator (Gooch and Housego) controlled by a microcontroller (Jaeger ADwin Pro II). For ns-pulses, we use in addition an in-fiber amplitude electro-optic modulator (aEOM) (Jenoptik AM635) controlled by an arbitrary wave generator (Tektronix AWG5014). For phase modulation, we use an in-fiber phase EOM (pEOM) (Jenoptik PM635). The EOMs are stabilized for maximum transmission. All laser paths are combined and focused on the emitter of interest. In the detection path the resonant and off-resonant laser light is filtered by a \unit[593]{nm} long-pass (Thorlabs, FELH0600) and a \unit[628]{nm} long-pass (Semrock, TLP01-628) filter. The resulting PSB photons are detected by an APD (Laser Components), which generates analog pulses that can be recorded by both a timetagger (Hydraharp, PicoQuant) and the microcontroller. The
setup is controlled and measurements are performed with
a PC and the python 3 framework QMI 0.37~\cite{raa_qmi_2023}.
\begin{figure}
	\includegraphics[width=0.8\linewidth]{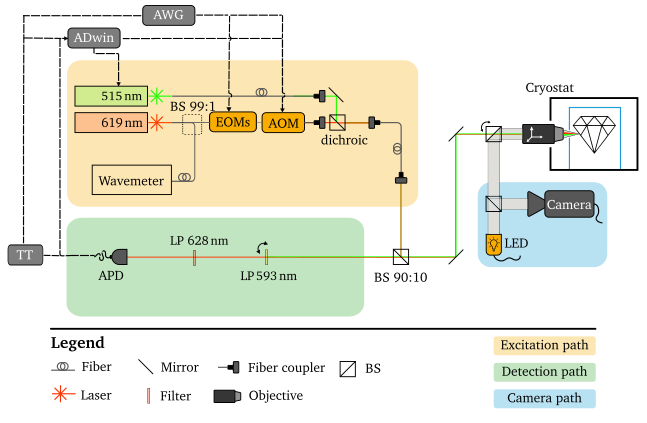}
	\caption{Confocal set-up  \label{fig:confocal set up}
 \label{fig:suppl_confocal_setup}}
\end{figure}

\section{Optical pulse characterization and optimization} \label{sec:Optical pulse characterization and optimization}
To create the optical $\uppi$-pulses, a \unit[26]{ns} AOM-pulse (\unit[25]{ns} rise time of the AOM and \unit[1]{ns} optical pulse) is applied, during which an aEOM suppresses the light during the ramping up and ramping down of the AOM pulse. During the window in which the aEOM is not suppressing a short \unit[1]{ns} resonant light pulse is transmitted. All pulses are generated by the AWG. \\
\\
The suppression of the light by the aEOM is optimized by sweeping the amplitude of the aEOM pulse and optimizing for the minimal number of photon counts detected during the ramping-up time of the AOM. An identical but reverse sign pulse is applied after the pulse sequence to prevent charging of the aEOM.\\
\\
The optical $\uppi$/2-pulses are characterized by fixing the amplitude of AOM wave, sweeping the amplitude of the pulse of the aEOM and, integrating the PSB-photons counts after the pulse. This results in a sinusoid response which we fit to determine the amplitude of the aEOM wave corresponding to a $\uppi$/2-pulse. \\
\\
For the Ramsey experiments, we noticed the amplitude of the second $\uppi$/2-pulse to vary with waiting time, $\uptau$, most likely to be caused by charging of the aEOM. We mitigate this effect by calibrating the amplitude of the second $\uppi$/2-pulse as a function of waiting time. We sweep the amplitude of the pulse of the aEOM and minimize the difference in detected photon counts between the first optimized pulse and the second $\uppi$/2-pulse.

\section{CRC Optical Ramsey Interferometry measurements} \label{sec:CRC Optical Ramsey Interferometry measurements}
Additional data to Fig.~5 of the main text. We sweep the CRC threshold, $C_{\text{pass}}$, while keeping $C_{\text{repump}}=10$. For $C_{\text{pass}}=1$, we implement $C_{\text{repump}}=1$, which is effectively equivalent to not implementing a CRC sequence but instead use a conditional repump, where we only repump once the emitter is in the dark state. 
\begin{figure}
	\includegraphics[width=\linewidth]{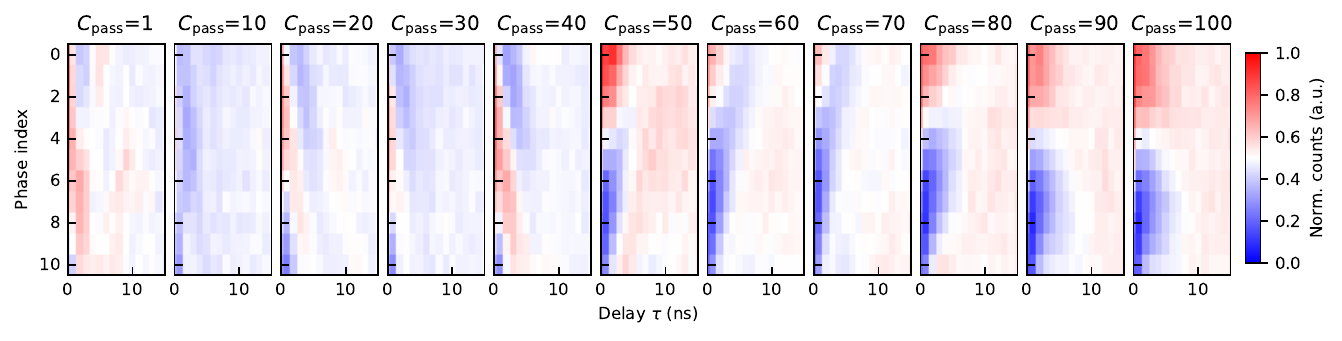}
	\caption{CRC Ramsey additional data, the phase index is based on the voltage step applied to the pEOM. Phase index 0 represents no phase difference between the 2 $\uppi$/2-pulses, which should result in maximum counts detected at $\uptau$=0. \label{fig:suppl_ramsey2d}}
\end{figure}

\section{CRC linescans} \label{sec:CRC linescans}
Additional data to Fig.~4 of the main text.
\begin{figure}[ht]
	\includegraphics[width=\linewidth]{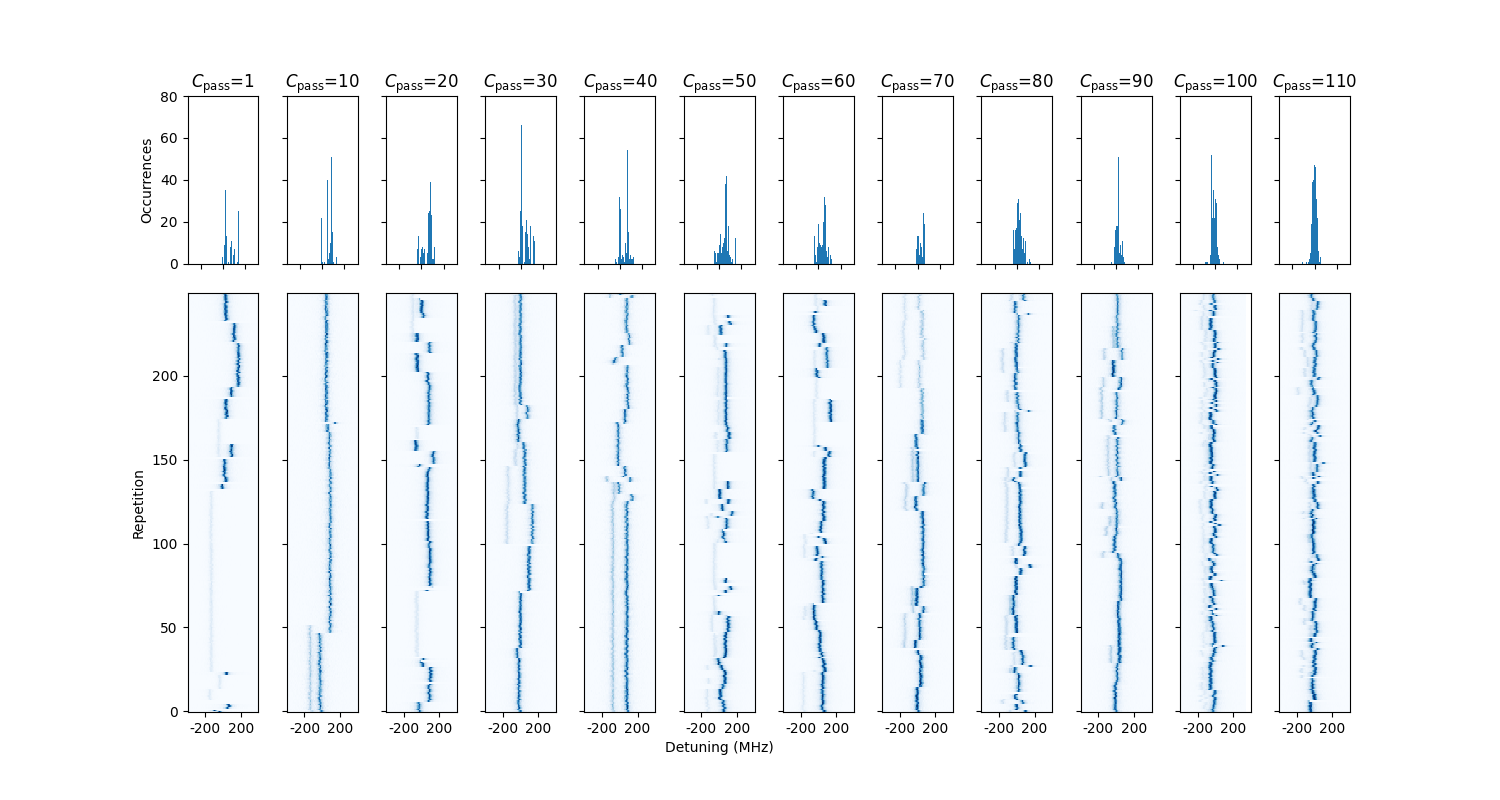}
	\caption{CRC linescan additional data of SnV$^-$A.  Bottom: The fluorescence of 250 PLE scans each taken in $\sim$\unit[1.3]{GHz/s} over the optical transition of SnV$^-$A, preceded by a CRC for various threshold $C_{\text{pass}}$ and a fixed threshold $C_{\text{repump}}=10$. We attribute the second red-detuned resonance detected to another nearby emitter. Top: The distribution of the fitted centers of the individual scans, filtered for the bright emitter.  \label{fig:suppl_cr_linescan}}
\end{figure}

\begin{figure}[ht]
	\includegraphics[width=\linewidth]{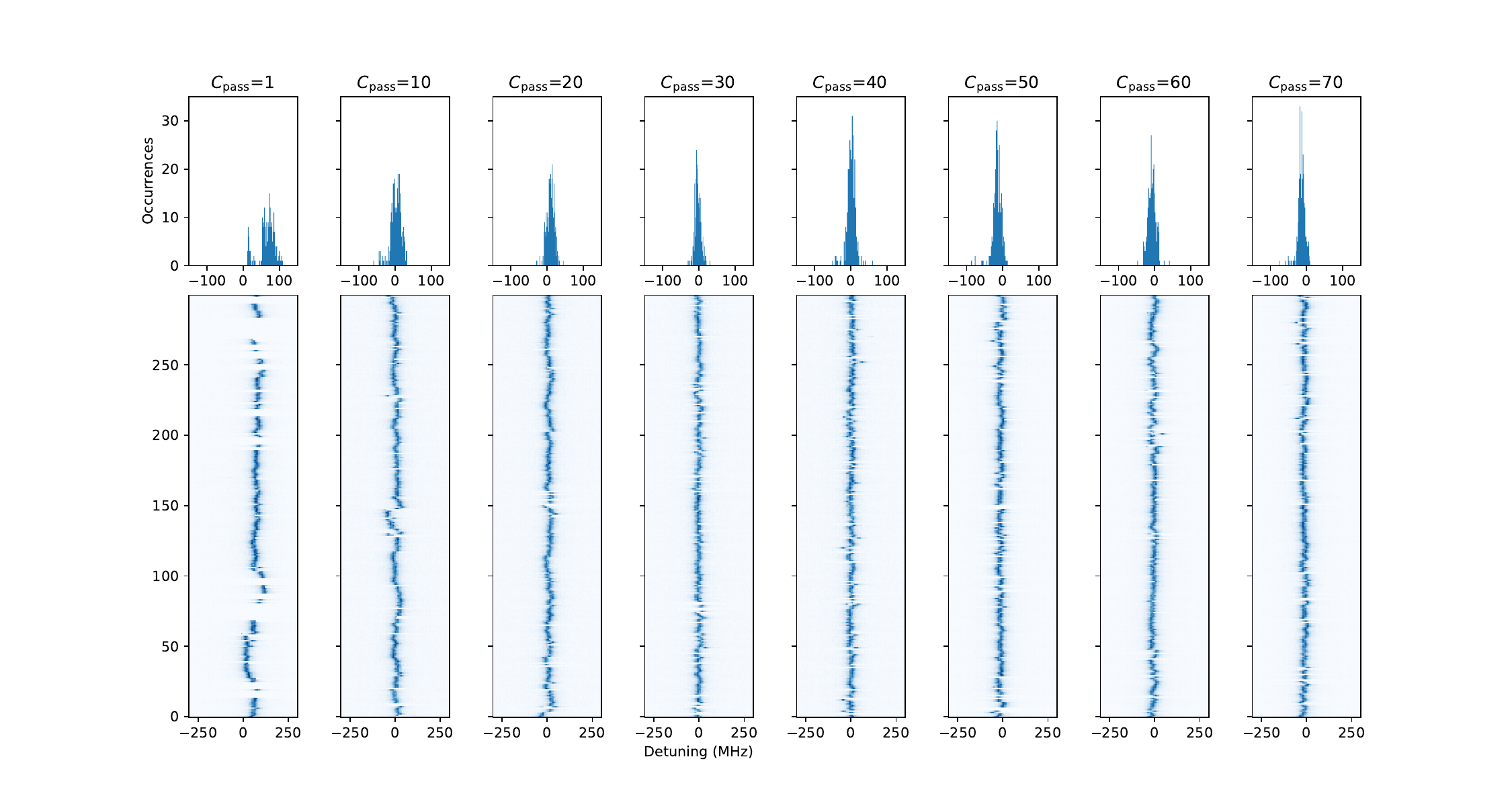}
	\caption{CRC linescan additional data of an in a waveguide embedded SnV$^-$B. Bottom: The fluorescence of 300 PLE scans each taken in $\sim$\unit[1.3]{GHz/s} over the optical transition of SnV$^-$A, preceded by a CRC for various threshold $C_{\text{pass}}$ and a fixed threshold $C_{\text{repump}}=10$. Top: The distribution of the fitted centers of the individual scans. \label{fig:suppl_cr_linescan_LTfiber1}}
\end{figure}
\section{Author contributions}
J.M.B. and L.D.S conducted the experiments and analysed the data. T.Y. performed the diamond overgrowth. J.M.B. prepared the diamond sample with SnV centers, SnV-A. T.T. helped with the measurements on the waveguide sample, SnV-B, which was fabricated by N.C. and M.P. designed and built the waveguide set-up. J.M.B., L.D.S., M.P., H.K.C.B., C.W. were involved in building parts of the set-up used for the main part of the measurements. J.M.B., L.D.S. and R.H. wrote the manuscript with input of all authors. R.H. and L.D.S. supervised the experiments.\\
\textbf{Data availability:} The data sets that support this manuscript are available at 4TU.ResearchData~\cite{brevoord_data_2023}.
\end{document}